\documentclass[12pt,superscriptaddress,aps,prd,preprint,showpacs]{revtex4}

\usepackage[dvips]{graphicx}
\usepackage{amsfonts}
\usepackage{slashed}
\usepackage[latin1]{inputenc}
\usepackage{amsmath}
\usepackage{hyperref}
\usepackage[svgnames]{xcolor}

\newcommand{\bea}{\begin{eqnarray}}
\newcommand{\eea}{\end{eqnarray}}

\begin{document}

\title{On perturbative aspects of a nonminimal Lorentz-violating QED with CPT-even dimension-5 terms}

\author{T. Mariz}
\affiliation{Instituto de F\'{\i}sica, Universidade Federal de Alagoas,\\ 57072-900, Macei\'o, Alagoas, Brazil}
\email{max.melo,tmariz@fis.ufal.br}

\author{M. Melo}
\affiliation{Instituto de F\'{\i}sica, Universidade Federal de Alagoas,\\ 57072-900, Macei\'o, Alagoas, Brazil}
\email{max.melo,tmariz@fis.ufal.br}

\author{J. R. Nascimento}
\affiliation{Departamento de F\'{\i}sica, Universidade Federal da Para\'{\i}ba,\\
 Caixa Postal 5008, 58051-970, Jo\~ao Pessoa, Para\'{\i}ba, Brazil}
\email{jroberto,petrov@fisica.ufpb.br}

\author{A. Yu. Petrov}
\affiliation{Departamento de F\'{\i}sica, Universidade Federal da Para\'{\i}ba,\\
 Caixa Postal 5008, 58051-970, Jo\~ao Pessoa, Para\'{\i}ba, Brazil}
\email{jroberto,petrov@fisica.ufpb.br}

\begin{abstract}
In this paper, we explicitly calculate the lower CPT-even one-loop quantum corrections in nonminimal Lorentz-violating spinor QED with all possible CPT-even dimension-5 operators. Within our calculations, we restrict ourselves to the cases when these parameters are completely expressed in terms of one constant vector. 
\end{abstract}


\maketitle

\section{Introduction}

Possibilities of breaking the Lorentz symmetry are actively studied now. The paradigmatic framework for describing the Lorentz-violating (LV) effects within the quantum field theory is the minimal Lorentz-violating Standard Model Extension (LV SME) originally formulated in \cite{ColKost1,ColKost2}. A detailed description of its electromagnetic sector is presented in \cite{KosPic}. Further, the interest in nonminimal LV extensions of QED, being motivated by studies of phenomenology of elementary particles where dealing with higher-dimensional operators plays an important role for a long time, beginning from the Fermi Lagrangian, emerged naturally. An important role was played by the paper \cite{KosMew}, where the simplest examples of such operators were introduced in the LV context for the first time. The first studies of their perturbative impact have been performed in \cite{aether,Mariz:2010fm,aether0,aether1,MNP}. Further, all possible LV extensions of SME with dimensions up to 6 for fermion-dependent operators were listed in \cite{KosLi}.

Therefore, the natural question consists in studying loop corrections generated by all these nonminimal operators, beginning from dimension 5 and so on. The lower quantum corrections generated by all possible CPT-odd dimension-5 additive terms have been obtained explicitly in our previous paper \cite{ourd5}, so the natural continuation of that study consists in obtaining the quantum corrections generated by all CPT-even ones with this dimension. Within this study, we restrict ourselves by the first, leading order of the nonminimal operators, taking in mind the aim to generate the CPT-even term similar to that one obtained in \cite{aether,aether0,aether1}, and calculate such a term for all dimension-5 CPT-even operators. For the sake of the simplicity, we consider the case when CPT-even constant LV tensors are completely described in terms of one LV vector, similarly to the scenario considered in \cite{ourd5} for CPT-odd dimension-5 operators, and in \cite{aether,aether0,aether1} for the aether term.

The structure of the paper looks as follows. In section 2, we define our model and calculate the effective action.  We perform the one-loop calculations in section 3 to study the generation of the aether term. Finally, in section 4, we discuss our results.

\section{Nonminimally extended Lorentz-violating QED}\label{model}

In this paper, we are interested in analyzing the following nonminimal LV extended QED Lagrangian with dimension-5 terms:
\begin{eqnarray}\label{start}
\mathcal{L}_{\psi} &=& \textstyle{1\over2}\bar{\psi}(i\slashed{D} -m)\psi -\textstyle{1\over2}m^{(5)\alpha\beta}\bar{\psi}iD_{(\alpha}iD_{\beta)}\psi -\textstyle{1\over2}im_5^{(5)\alpha\beta}\bar{\psi}\gamma_5iD_{(\alpha}iD_{\beta)}\psi \nonumber\\
&&-\textstyle{1\over2}a^{(5)\mu \alpha \beta}\bar{\psi}\gamma_{\mu}iD_{(\alpha}iD_{\beta)}\psi -\textstyle{1\over2}b^{(5)\mu \alpha \beta}\bar{\psi}\gamma_5 \gamma_{\mu}iD_{(\alpha}iD_{\beta)}\psi \nonumber\\
&&-\textstyle{1\over4}H^{(5)\mu\nu\alpha\beta}\bar{\psi}\sigma_{\mu\nu}iD_{(\alpha}iD_{\beta)}\psi +\mathrm{h.c.} -\textstyle{1\over2}m^{(5)\alpha\beta}_F\bar{\psi}F_{\alpha\beta}\psi -\textstyle{1\over2}im_{5F}^{(5)\alpha\beta}\bar{\psi}\gamma_5F_{\alpha\beta}\psi \nonumber\\
&&-\textstyle{1\over2}a_F^{(5)\mu \alpha \beta} \bar{\psi}\gamma_\mu F_{\alpha \beta} \psi -\textstyle{1\over2}b_F^{(5)\mu \alpha \beta} \bar{\psi}\gamma_5 \gamma_\mu F_{\alpha \beta} \psi -\textstyle{1\over4}H^{(5)\mu\nu\alpha\beta}_F\bar{\psi}\sigma_{\mu\nu}F_{\alpha\beta}\psi,
\end{eqnarray}
where $F_{\alpha \beta} = \partial_\alpha A_\beta - \partial_\beta A_\alpha $, $D_\mu\psi= \partial_\mu\psi+ie A_\mu\psi$, and
\begin{eqnarray}
iD_{(\alpha} i D_{\beta)}\psi&=&\textstyle{1\over2}\left(iD_\alpha i D_\beta + iD_\beta iD_\alpha\right)\psi\nonumber \\
&=&-\partial_\alpha \partial_\beta\psi -ie[A_\beta \partial_\alpha+A_\alpha \partial_\beta +\textstyle{1\over2}(\partial_\alpha A_\beta + \partial_\beta A_\alpha)]\psi+e^2A_\alpha A_\beta\psi.
\end{eqnarray}
The expression (\ref{start}) includes all dimension-5 LV couplings defined in \cite{KosLi}. In fact, our aim will consist of studying the CPT-even contributions to the one-loop effective action of the gauge field, e.g., the aether term. It is clear that, in the first order, exclusively the terms involving even-rank constant tensors must be considered since only contractions of such tensors with the Minkowski metric and the Levi-Civita symbol, in four-dimensional space-time, could yield a CPT-even contribution. Therefore, within this study, we deal with all possible CPT-even dimension-5 couplings. This allows us to reduce our Lagrangian to
\begin{eqnarray}\label{start2}
\mathcal{L}_{\psi} &=& \textstyle{1\over2}\bar{\psi}(i\slashed{D} -m)\psi -\textstyle{1\over2}m^{(5)\alpha\beta}\bar{\psi}iD_{(\alpha}iD_{\beta)}\psi -\textstyle{1\over2}im_5^{(5)\alpha\beta}\bar{\psi}\gamma_5iD_{(\alpha}iD_{\beta)}\psi \nonumber\\
&&-\textstyle{1\over4}H^{(5)\mu\nu\alpha\beta}\bar{\psi}\sigma_{\mu\nu}iD_{(\alpha}iD_{\beta)}\psi +\mathrm{h.c.} -\textstyle{1\over2}m_F^{(5)\alpha\beta}\bar{\psi}F_{\alpha\beta}\psi -\textstyle{1\over2}im_{5F}^{(5)\alpha\beta}\bar{\psi}\gamma_5F_{\alpha\beta}\psi \nonumber\\
&&-\textstyle{1\over4}H_F^{(5)\mu\nu\alpha\beta}\bar{\psi}\sigma_{\mu\nu}F_{\alpha\beta}\psi.
\end{eqnarray}
We can also rewrite the expression (\ref{start2}) as follows:
\begin{eqnarray}
\mathcal{L}_{\psi} &=& \bar{\psi} (i\slashed{\partial} +T^{\alpha\beta} \partial_\alpha \partial_\beta-m -e\slashed{A} +ieT^{\alpha\beta} \nabla_\alpha A_\beta -eT_F^{\alpha\beta} (\partial_\alpha A_\beta) -e^2T^{\alpha\beta}A_\alpha A_\beta)\psi,
\end{eqnarray}
where
\begin{equation}\label{T}
T^{\alpha\beta} = m^{(5)\alpha \beta}+im_5^{(5)\alpha \beta} \gamma_5+\textstyle{1\over2}H^{(5)\mu\nu\alpha\beta}\sigma_{\mu\nu},
\end{equation}
\begin{equation}\label{TF}
T_F^{\alpha\beta} = m_F^{(5)\alpha \beta}+im_{F5}^{(5)\alpha \beta} \gamma_5+\textstyle{1\over2}H_F^{(5)\mu\nu\alpha\beta}\sigma_{\mu\nu},
\end{equation}
and we introduced the definition $\nabla_\alpha A_\beta\equiv 2A_\beta\partial_\alpha +(\partial_\alpha A_\beta)$.

Then, the corresponding effective action, after the integration over the spinor fields, is given by
\begin{eqnarray}\label{Seff}
S_\mathrm{eff} = -i\mathrm{Tr}\ln(\slashed{p} -T^{\alpha\beta}p_\alpha p_\beta-m -e\slashed{A} +eT^{\alpha\beta} \nabla_\alpha(p,k) A_\beta +ieT_F^{\alpha\beta} k_\alpha A_\beta -e^2T^{\alpha\beta}A_\alpha A_\beta), 
\end{eqnarray}
with, in the momentum space, $\nabla_\alpha(p,k)=2p_\alpha+k_\alpha$, $i\partial_\alpha \psi=p_\alpha \psi$, and $i\partial_\alpha A_\beta =k_\alpha A_\beta$. Here, $\mathrm{Tr}$ stands for the trace over the Dirac matrices, as well as the trace over the integration in momentum and coordinate spaces.

We can expand Eq.~(\ref{Seff}) in power series in external fields as 
\begin{equation}
S_\mathrm{eff}=S_\mathrm{eff}^{(0)}+\sum_{n=1}^\infty S_\mathrm{eff}^{(n)},
\end{equation} 
where $S_\mathrm{eff}^{(0)}=-i\mathrm{Tr}\ln G^{-1}(p)$ and 
\begin{eqnarray}\label{Seffn}
S_\mathrm{eff}^{(n)} &=& \frac{i}{n}\mathrm{Tr}[G(p)(e\slashed{A} -eT^{\alpha\beta} \nabla_\alpha(p,k) A_\beta -ieT_F^{\alpha\beta} k_\alpha A_\beta +e^2T^{\alpha\beta}A_\alpha A_\beta)]^n,
\end{eqnarray}
with 
\begin{equation}
G(p)=\frac1{\slashed{p} -T^{\alpha\beta}p_\alpha p_\beta -m}.
\end{equation}
Since, for this step, we are interested in the induced CPT-even terms, we need to work only with terms of first order in (\ref{T}) and (\ref{TF}) and second order in $A^\mu$.

\begin{figure}[h]
\centering{}\includegraphics[scale=0.9]{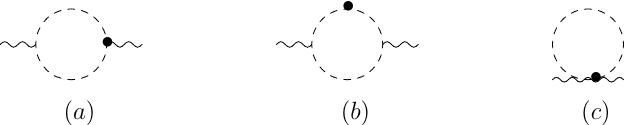} \caption{General form of first order Lorentz-breaking contributions.}
\label{Fig1}
\end{figure}

First, we write down the contributions from quartic vertices given by Fig.~\ref{Fig1}c, i.e., for $n=1$ in (\ref{Seffn}). Their contribution to the effective action  is 
\begin{equation}\label{S1}
S^{(1)}_\mathrm{eff}=i\int d^4x \Pi_1^{\mu \nu}A_\mu A_\nu,
\end{equation} with
\begin{equation}\label{Pi1}
\Pi^{\mu\nu}_1= e^2\int \frac{d^4p}{(2\pi)^4}\mathrm{tr}\,S(p)T^{\mu\nu},
\end{equation}
where we have considered 
\begin{equation}
G(p)=S(p)+S(p) T^{\alpha\beta}p_\alpha p_\beta S(p)+\cdots.
\end{equation}
We note that since the vertex is already of the first order in a corresponding LV parameter, the propagator is not needed to be modified now, i.e., we can use $S(p)=(\slashed{p}-m)^{-1}$.

\vspace*{3mm}	
\vspace*{3mm}
	
It is immediate to see that the contributions proportional to $m_5^{(5)\alpha\beta}$ and $H^{(5)\mu\nu\alpha\beta}$ from~(\ref{Pi1}) for this Feynman diagram identically vanish since corresponding traces are equal to zero. Then, 
\begin{equation}
\Pi^{\mu\nu}_{1,m_5}=\Pi^{\mu\nu}_{1,H}=0.
\end{equation}
Thus, the only nontrivial contribution is proportional to $m^{(5)\alpha\beta}$, which can be evaluated by taking into account 
\begin{equation}
\label{expr}
m^{(5)\alpha\beta}=\mathsf{m}^\alpha \mathsf{m}^\beta -\zeta{{\mathsf{m}^2}\over4}g^{\alpha\beta}. 
\end{equation}
It yields
\begin{equation}\label{Pi12}
\Pi^{\mu\nu}_{1,m}= -\frac{i e^2 m^3}{8 \pi ^2 \epsilon'}\left(\zeta  \mathsf{m}^2 g^{\mu  \nu }-4 \mathsf{m}^{\mu } \mathsf{m}^{\nu }\right) -\frac{i e^2 m^3}{16 \pi ^2}\left(\zeta  \mathsf{m}^2 g^{\mu  \nu }-4 \mathsf{m}^{\mu } \mathsf{m}^{\nu }\right),
\end{equation}
where $\frac1{\epsilon'}=\frac1{\epsilon}-\ln\frac{m}{\mu'}$, with $\epsilon=4-D$ and $\mu'^2=4\pi\mu^2e^{-\gamma}$. It is important to note that when $\zeta=1$, $m^{(5)\alpha\beta}$ becomes traceless. This is a significant point that requires further exploration.

\begin{figure}[h]
\centering{}\includegraphics[scale=0.9]{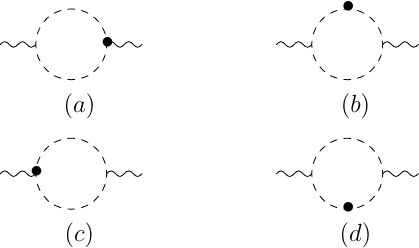} \caption{General form of first order Lorentz-breaking contributions} 
\label{Fig2}
\end{figure}

The remaining LV contributions, which can contribute to graphs with triple vertices in Fig.~\ref{Fig1}, i.e., (a) and (b), for $n=2$ in (\ref{Seffn}), are given by the Feynman diagram depicted in Fig.~\ref{Fig2}.

\vspace*{3mm}
\vspace*{3mm}

Then, the effective action can be written as
\begin{equation}\label{S2}
S^{(2)}_\mathrm{eff}=\textstyle{i\over2}\int d^4x (\Pi_{2a}^{\mu\nu}+\Pi_{2b}^{\mu\nu}+\Pi_{2c}^{\mu\nu}+\Pi_{2d}^{\mu\nu})A_\mu A_\nu,
\end{equation}
with
\begin{subequations}\label{Pi2}
\begin{eqnarray}
\label{Pi2a}\Pi^{\mu \nu}_{2a}&=&-e^2 \int \frac{d^4p}{(2\pi)^4}\mathrm{tr}\,S(p) (T^{\kappa\mu} \nabla_\kappa(p,k)+iT_F^{\kappa\mu}k_\kappa)S(p-k) \gamma^\nu,\\
\label{Pi2b}\Pi^{\mu \nu}_{2b}&=&-e^2\int \frac{d^4p}{(2\pi)^4}\mathrm{tr}\,S(p) \gamma^\mu S(p-k)(T^{\kappa\nu}\nabla_\kappa(p-k,-k)+iT_F^{\kappa\nu}(-k_\kappa)),\\
\label{Pi2c}\Pi^{\mu \nu}_{2c}&=&e^2\int \frac{d^4p}{(2\pi)^4}\mathrm{tr} S(p) T^{\kappa\lambda}p_\kappa p_\lambda S(p) \gamma^\mu S(p-k)\gamma^\nu, \\
\label{Pi2d}\Pi^{\mu \nu}_{2d}&=&e^2\int \frac{d^4p}{(2\pi)^4}\mathrm{tr} S(p) \gamma^\mu S(p-k) T^{\kappa\lambda}(p-k)_\kappa (p-k)_\lambda S(p-k)\gamma^\nu.
\end{eqnarray}
\end{subequations}

Except of the $H^{(5)\mu\nu\alpha\beta}$ and $H^{(5)\mu\nu\alpha\beta}_F$ parameters, all other LV parameters are given by second-rank tensors. We note again that our aim consists in obtaining the CPT-even contribution to the effective action given by the term
\bea
{\cal L}_{even}=-\textstyle{1\over4}(k_F)_{\mu\nu\lambda\rho}F^{\mu\nu}F^{\lambda\rho}.
\eea
If we use the second-rank constant tensors (let us denote them as $c_{\mu\nu}$), we must have
\bea
(k_F)_{\mu\nu\lambda\rho}\propto g_{\mu\lambda}c_{\nu\rho} -g_{\mu\rho}c_{\nu\lambda} +g_{\nu\rho}c_{\mu\lambda} -g_{\nu\lambda}c_{\mu\rho},
\eea
so, ${\cal L}_{even}\propto c_{\mu\nu}F^{\mu\lambda}F^{\nu}_{\phantom{\nu}\lambda}$. It is clear that this term vanishes identically if $c_{\mu\nu}$ is antisymmetric. Thus, the terms  proportional to $m_{F}^{(5)\alpha\beta}$ and $m_{F5}^{(5)\alpha\beta}$ in (\ref{Pi2a}) and (\ref{Pi2b}) will not yield a nontrivial contribution to the desired term. In fact, for $m_{F}^{(5)\alpha\beta}$, 
\begin{equation}
{\cal L}_{m_F}=(\textstyle{i\over2}\Pi^{\mu \nu}_{2a,m_F}+\textstyle{i\over2}\Pi^{\mu \nu}_{2b,m_F})A_\mu A_\nu=0,
\end{equation}
after the calculation of the integrals, and, for $m_{F5}^{(5)\alpha\beta}$, 
\begin{equation}
{\cal L}_{m_{F5}}=(\textstyle{i\over2}\Pi^{\mu \nu}_{2a,m_{F5}}+\textstyle{i\over2}\Pi^{\mu \nu}_{2b,m_{F5}})A_\mu A_\nu=0,
\end{equation}
after the calculation of the traces.




Now, we need to calculate the remaining insertions of Eq.~(\ref{Pi2}). Initially, let us analyze the coefficient $H^{(5)\mu\nu\alpha\beta}_F$, whose contributions are present in both Eqs.~(\ref{Pi2a}) and (\ref{Pi2b}). The results are
\begin{eqnarray}\label{LHF}
{\cal L}_{H_F}&=&(\textstyle{i\over2}\Pi^{\mu \nu}_{2a,H_F}+\textstyle{i\over2}\Pi^{\mu \nu}_{2b,H_F})A_\mu A_\nu \\
&=&\left[-\frac{e^2 m}{2 \pi ^2 \epsilon'} -\frac{e^2 m}{2\pi^2} \left(1-\sqrt{\frac{4 m^2}{k^2}-1} \cot^{-1}\left(\sqrt{\frac{4 m^2}{k^2}-1}\right)\right)\right]H_F^{(5)\kappa\mu\lambda\nu}k_\kappa k_\lambda A_\mu A_\nu, \nonumber
\end{eqnarray}
where
\begin{eqnarray}\label{HF}
H_F^{(5)\kappa\mu\lambda\nu}k_\kappa k_\lambda &=& \frac12\left(g^{\mu  \nu } \left(2 (H_F\cdot k)^2-\zeta  H_F^2 k^2\right)+H_F^{\mu } \left(2 k^2 H_F^{\nu }-2 k^{\nu } (H_F\cdot k)\right)\right. \nonumber\\
&&\left.+k^{\mu } \left(\zeta  H_F^2 k^{\nu }-2 H_F^{\nu } (H_F\cdot k)\right)\right),
\end{eqnarray}
in which our fourth-rank tensor is chosen to be described in terms of the unique vector similar to (\ref{expr}), i.e., by the expression
\begin{equation}
H_F^{(5)\kappa\mu\lambda\nu} = g^{\kappa\lambda}H_F^{\mu\nu} -g^{\kappa\nu}H_F^{\mu\lambda} +g^{\mu\nu}H_F^{\kappa\lambda} -g^{\mu\lambda}H_F^{\kappa\nu},
\end{equation}
with
\begin{equation}
H_F^{\mu\nu}=H_F^\mu H_F^\nu -\zeta{{H_F^2}\over4}g^{\mu\nu}. 
\end{equation}
In the case where $\zeta=1$, the coefficient $H_F^{\mu\nu}$ is traceless, leading to the expected outcome of $H_F^{(5)\kappa\mu\lambda\nu}$ being double traceless.

Actually, the divergent part of (\ref{LHF}), denoted as 
\begin{equation}
{\cal L}_{H_F,div}=-\frac{e^2m}{8\pi^2\epsilon'}H_F^{(5)\kappa\mu\lambda\nu}F_{\kappa\mu}F_{\lambda\nu},
\end{equation}
has already been calculated in a previous study performed in \cite{Maluf2013}. This term also has been discussed in \cite{Carvalho}, where a slightly different coupling (with $\sigma_{\mu\nu}\gamma_5$ matrix instead of $\sigma_{\mu\nu}$ one was used) has been considered. In the limit of $\frac{k^2}{m^2}\ll0$, the finite part of the Lagrangian (\ref{LHF}) takes the form
\begin{equation}
{\cal L}_{H_F,fin}=-\frac{e^2}{96 \pi ^2 m}H_F^{(5)\kappa\mu\lambda\nu}F_{\kappa\mu}\Box F_{\lambda\nu}.
\end{equation}

For other dimension-5 operators, those one proportional to $m^{(5)\alpha \beta}$, $m^{(5)\alpha \beta}_5$, and $H^{(5)\mu\nu\alpha\beta}$, we must take into account all the expressions (\ref{Pi2}), i.e., the graphs $(a)$, $(b)$, $(c)$, and $(d)$ of Fig.~(\ref{Fig2}).

Then, it is easy to see that if we consider the $m^{(5)\alpha \beta}_5$ insertion, the graphs (\ref{Pi2a}) and (\ref{Pi2b}) vanish being proportional to ${\rm tr}\,(\slashed{p}+m)\gamma_5(\slashed{p}-\slashed{k}+m)\gamma^{\nu}=0$ and ${\rm tr}\,(\slashed{p}+m)\gamma^\mu(\slashed{p}-\slashed{k}+m)\gamma_5=0$. The graphs (\ref{Pi2c}) and (\ref{Pi2d}) yield contributions proportional to ${\rm tr}\,(\slashed{p}+m)\gamma_5(\slashed{p}+m)\gamma^{\mu}(\slashed{p}-\slashed{k}+m)\gamma^\nu=-{\rm tr}\,(p^2-m^2)\gamma_5\gamma^{\mu}(\slashed{p}-\slashed{k}+m)\gamma^\nu$ and ${\rm tr}\,(\slashed{p}+m)\gamma^\mu(\slashed{p}-\slashed{k}+m)\gamma_5(\slashed{p}-\slashed{k}+m)\gamma^\nu=-{\rm tr}\,((p-k)^2-m^2)(\slashed{p}+m)\gamma^\mu\gamma_5\gamma^\nu$, which also vanish. Thus, we have
\begin{equation}
{\cal L}_{m_{5}}=(i\Pi^{\mu\nu}_{1,m_5}+\textstyle{i\over2}\Pi^{\mu \nu}_{2a,m_{5}}+\textstyle{i\over2}\Pi^{\mu \nu}_{2b,m_{5}}+\textstyle{i\over2}\Pi^{\mu \nu}_{2c,m_{5}}+\textstyle{i\over2}\Pi^{\mu \nu}_{2d,m_{5}})A_\mu A_\nu=0.
\end{equation}

So, we rest with the insertions $m^{(5)\alpha \beta}$ and $H^{(5)\mu\nu\alpha\beta}$. First, we find the contribution with $H^{(5)\mu\nu\alpha\beta}$, which is given by
\begin{eqnarray}\label{LH}
{\cal L}_H&=&(i\Pi^{\mu\nu}_{1,H}+\textstyle{i\over2}\Pi^{\mu \nu}_{2a,H}+\textstyle{i\over2}\Pi^{\mu \nu}_{2b,H}+\textstyle{i\over2}\Pi^{\mu \nu}_{2c,H}+\textstyle{i\over2}\Pi^{\mu \nu}_{2d,H})A_\mu A_\nu \\
&=&\left[\frac{e^2 m}{\pi ^2 \epsilon'} +\frac{e^2 m}{\pi^2} \left(1-\sqrt{\frac{4 m^2}{k^2}-1} \cot^{-1}\left(\sqrt{\frac{4 m^2}{k^2}-1}\right)\right)\right]H^{(5)\kappa\mu\lambda\nu}k_\kappa k_\lambda A_\mu A_\nu, \nonumber
\end{eqnarray}
with, as in (\ref{HF}), 
\begin{eqnarray}
H^{(5)\kappa\mu\lambda\nu}k_\kappa k_\lambda &=& \frac12\left(g^{\mu  \nu } \left(2 (H\cdot k)^2-\zeta  H^2 k^2\right)+H^{\mu } \left(2 k^2 H^{\nu }-2 k^{\nu } (H\cdot k)\right)\right. \nonumber\\
&&\left.+k^{\mu } \left(\zeta  H^2 k^{\nu }-2 H^{\nu } (H\cdot k)\right)\right),
\end{eqnarray}
where we again have done the choice allowing us to describe our fourth-rank tensor by a single vector, namely,
\begin{equation}
H^{(5)\kappa\mu\lambda\nu} = g^{\kappa\lambda}H^{\mu\nu} -g^{\kappa\nu}H^{\mu\lambda} +g^{\mu\nu}H^{\kappa\lambda} -g^{\mu\lambda}H^{\kappa\nu},
\end{equation}
and, similarly to (\ref{expr}), we required
\begin{equation}
H^{\mu\nu}=H^\mu H^\nu -\zeta{{H^2}\over4}g^{\mu\nu}, 
\end{equation}
which, as expected, we have a double traceless coefficient, i.e., ${H^{(5)\kappa\mu}}_{\kappa\mu}=0$, when $\zeta=1$. 
It is easy to check that $H^{(5)\kappa\mu\lambda\nu}k_\kappa k_\lambda k_{\mu}=H^{(5)\kappa\mu\lambda\nu}k_\kappa k_\lambda k_{\nu}=0$ which guarantees the transversality of the operator $H^{(5)\kappa\mu\lambda\nu}k_\kappa k_\lambda$ and hence the gauge symmetry of our result (\ref{LH}).
The divergent part of Eq.~(\ref{LH}) can be rewritten as 
\begin{equation}
{\cal L}_{H,div}=\frac{e^2m}{4\pi^2\epsilon'}H^{(5)\kappa\mu\lambda\nu}F_{\kappa\mu}F_{\lambda\nu}.
\end{equation}
Additionally, if we take the limit of $\frac{k^2}{m^2}\ll0$, we can express the finite part as 
\begin{equation}
{\cal L}_{H,fin}=\frac{e^2}{48 \pi ^2 m}H^{(5)\kappa\mu\lambda\nu}F_{\kappa\mu}\Box F_{\lambda\nu}.
\end{equation}
It is interesting to note that for the special relation of our LV parameters, namely, $H^\mu=\textstyle{1\over2}H_F^\mu$, the one-loop corrections (\ref{LHF}) and (\ref{LH}) mutually cancel. The similar effect occurs also in LV QED with CPT-odd dimension-5 parameters \cite{ourd5} and, in principle, can be treated as one more example of similarity of perturbative corrections generated by different LV parameters, analogous to the scenario discussed in \cite{Alts} for the minimal LV QED.

Now, it remains to consider $m^{(5)\alpha\beta}$, which we assume to have the form given by (\ref{expr}). In this case, we have contributions generated both by Fig.~\ref{Fig1}c and Fig.~\ref{Fig2}. Then, we obtain
\begin{eqnarray}\label{m5}
{\cal L}_{m}&=&(i\Pi^{\mu\nu}_{1,m}+\textstyle{i\over2}\Pi^{\mu \nu}_{2a,m}+\textstyle{i\over2}\Pi^{\mu \nu}_{2b,m}+\textstyle{i\over2}\Pi^{\mu \nu}_{2c,m}+\textstyle{i\over2}\Pi^{\mu \nu}_{2d,m})A_\mu A_\nu \nonumber \\
&=& {\cal L}_{m,div} +{\cal L}_{m,fin},
\end{eqnarray}
where
\begin{eqnarray}\label{resdiv}
{\cal L}_{m,div} &=& -\frac{e^2 m}{24 \pi ^2 \epsilon'}\left((3 \zeta -2) \mathsf{m}^2 \left(k^{\mu } k^{\nu }-k^2 g^{\mu  \nu }\right)+2 g^{\mu  \nu } (k\cdot \mathsf{m})^2+2 k^2 \mathsf{m}^{\mu } \mathsf{m}^{\nu } \right. \nonumber\\
&&\left. -2 (k\cdot \mathsf{m}) \left(k^{\nu } \mathsf{m}^{\mu }+k^{\mu } \mathsf{m}^{\nu }\right)\right) A_\mu A_\nu,
\end{eqnarray}
or, returning to the coordinate space, we can rewrite the divergent contribution as
\begin{eqnarray}\label{resdiv2}
{\cal L}_{m,div} &=& \frac{e^2 m(3 \zeta -2)}{48 \pi ^2 \epsilon'}\mathsf{m}^2F_{\mu\nu}F^{\mu\nu} -\frac{e^2 m}{12 \pi ^2 \epsilon'}\mathsf{m}^{\mu}F_{\mu\nu}\mathsf{m}_{\lambda}F^{\lambda\nu},
\end{eqnarray}
i.e. we have Maxwell-like and aether-like divergent terms. We note that if we introduce the traceless LV tensor $\tilde{m}^{(5)\mu\nu}=\mathsf{m}^{\mu}\mathsf{m}^{\nu}-\frac{1}{4}\mathsf{m}^2g^{\mu\nu}$ and choose $\zeta=1$ in (\ref{resdiv2}), the divergent result (\ref{resdiv}) will take the form
\begin{eqnarray}\label{resdiv3}
{\cal L}_{m,div} &=& -\frac{e^2 m}{12 \pi ^2 \epsilon'} {\tilde m}^{(5)\mu\lambda} F_{\mu\nu}{F_\lambda}^{\nu},
\end{eqnarray}
i.e. in this case the trace part of the tensor $\mathsf{m}^{\mu}\mathsf{m}^{\nu}$ is completely ruled out, and the result is completely described by its traceless part.

The finite part of the CPT-even contribution of the Lagrangian (\ref{m5}) is 
\begin{eqnarray}
\label{finpart}
{\cal L}_{m,fin} &=& (A\, \mathsf{m}^2 (k^\mu k^\nu-g^{\mu  \nu }k^2) +B g^{\mu\nu}(k\cdot \mathsf{m})^2 +C \frac{k^{\mu} k^{\nu}}{k^2}(k\cdot \mathsf{m})^2 +D k^2\mathsf{m}^{\mu} \mathsf{m}^{\nu} \nonumber\\
&& -D (k\cdot \mathsf{m}) (k^{\nu } \mathsf{m}^{\mu }+k^{\mu } \mathsf{m}^{\nu }))A_\mu A_\nu,
\end{eqnarray}
with
\begin{eqnarray}
A &=& -\frac{e^2 m}{144 \pi ^2 \sqrt{k^6 \left(4 m^2-k^2\right)}}\left[((21 \zeta -10)k^2+6 (3 \zeta -4) m^2) \sqrt{k^2(4 m^2-k^2)} \right. \nonumber\\
&&\left. +(24 (1-3 \zeta ) k^2 m^2 -24(3 \zeta -4) m^4 +6 (3 \zeta -2) k^4) \cot^{-1}\left(\sqrt{\frac{4 m^2}{k^2}-1}\right)\right],
\end{eqnarray}
\begin{eqnarray}
B &=& -\frac{e^2 m}{36 \pi ^2 \sqrt{k^6(4 m^2-k^2)}} \left[(7k^2+6m^2) \sqrt{k^2(4 m^2-k^2)} \right. \nonumber\\
&&\left. -3\left(8 k^2 m^2-k^4+8 m^4\right) \cot^{-1}\left(\sqrt{\frac{4 m^2}{k^2}-1}\right)\right],
\end{eqnarray}
\begin{eqnarray}
C &=& \frac{e^2 m}{12 \pi ^2 \sqrt{k^6(4 m^2-k^2)}}\left[(k^2+6 m^2) \sqrt{k^2(4 m^2-k^2)}\right. \nonumber\\
&&\left. -24 m^4 \cot^{-1}\left(\sqrt{\frac{4 m^2}{k^2}-1}\right)\right],
\end{eqnarray}
\begin{eqnarray}
D &=& -\frac{e^2 m}{36 \pi ^2 \sqrt{k^6 \left(4 m^2-k^2\right)}}\left[(4 k^2-12m^2) \sqrt{k^2(4 m^2-k^2)}\right. \nonumber\\
&&\left. +3 (k^2-4 m^2)^2 \cot^{-1}\left(\sqrt{\frac{4 m^2}{k^2}-1}\right)\right].
\end{eqnarray}
It is worth noting that the above Lagrangian ${\cal L}_m$ is invariant under gauge transformations, which requires that $D=B+C$. This identity is checked straightforwardly. In the IR limit of $\frac{k^2}{m^2}\ll0$, we find
\begin{eqnarray}
\label{ir}
A=-\frac{e^2(3\zeta-4)k^2}{480\pi^2m},\ B=\frac{e^2k^2}{80\pi^2m},\ C=-\frac{e^2k^2}{60\pi^2m},\ D=-\frac{e^2k^2}{240\pi^2m}. 
\end{eqnarray}
After implementing a Fourier transform and removing the coefficient $B$, it is possible to simplify the finite contribution (\ref{finpart}) into the form
\begin{eqnarray}
{\cal L}_{m,fin} = -\frac{1}{2}\mathsf{m}^2F^{\mu\nu}A(-\Box)F_{\mu\nu} -\frac{1}{2}F^{\mu\nu}\frac{(\mathsf{m}\cdot\partial)^2}{\Box}C(-\Box)F_{\mu\nu} +\mathsf{m}^{\mu}F_{\mu\nu}(D(-\Box))\mathsf{m}_{\lambda}F^{\lambda\nu},
\end{eqnarray}
where $A(-\Box)$, $C(-\Box)$, and $D(-\Box)$ are Fourier transforms for $A(k^2)$, $C(k^2)$, and $D(k^2)$. 
Rewriting this expression in terms of the traceless tensor $\tilde{m}^{(5)\mu\nu}=\mathsf{m}^{\mu}\mathsf{m}^{\nu}-\frac{1}{4}\mathsf{m}^2g^{\mu\nu}$, we find that, at $\zeta=1$, this expression takes the form
	\begin{eqnarray}
		{\cal L}_{m,fin} &=& -\frac{1}{2}F^{\mu\nu}\frac{\tilde{m}^{(5)\alpha\beta}\partial_{\alpha}\partial_{\beta}}{\Box}C(-\Box)F_{\mu\nu} +\tilde{m}^{(5)\mu\lambda}F_{\mu\nu}D(-\Box){F_\lambda}^{\nu},
	\end{eqnarray}
i.e. in this case the finite part is completely described by the traceless LV tensor $\tilde{m}^{(5)\mu\nu}$, and Lorentz symmetric terms are ruled out.

Taking into account (\ref{ir}), we see that in the IR limit we arrive at the finite part of our result given by the linear combination of four-derivative CPT-even terms $\tilde{m}^{(5)\alpha\beta}F^{\mu\nu}\partial_{\alpha}\partial_{\beta}F_{\mu\nu}$ and $\tilde{m}^{(5)\mu\lambda}F_{\mu\nu}\Box {F_\lambda}^\nu$. Explicitly, the one-loop contribution to the effective Lagrangian takes the form
\begin{eqnarray}
{\cal L}_{m,fin} &=& -\frac{e^2}{120\pi^2m}\tilde{m}^{(5)\alpha\beta}F^{\mu\nu}\partial_{\alpha}\partial_{\beta}F_{\mu\nu} +\frac{e^2}{240\pi^2m}\tilde{m}^{(5)\mu\lambda}F_{\mu\nu}\Box {F_\lambda}^\nu. \nonumber\\
\end{eqnarray}

We see that, for this LV additive term, terms with only two derivatives are canceled out in the finite part (actually, since $\mu$ dependence is related to the divergent part, we can conclude that absence of the second-derivative terms in the finite part is a consequence of our renormalization prescription). Thus, the first nontrivial contributions are the four-derivative ones. 

\section{Summary}

We calculated the CPT-even term for the nonminimal LV extended QED with all possible CPT-even dimension-5 operators, with the corresponding LV constant tensors are chosen to be expressed in terms of one LV vector. It turns out to be that in this case, unlike the schemes given in \cite{aether,aether1}, for all cases where a CPT-even contribution differs from zero, it is divergent. Nevertheless, it is a rather reasonable result explained by the non-renormalizability of our theory. Also, we found the finite part of the CPT-even term which involves fourth derivatives. Actually this is the second example of calculating four-derivative LV terms while the first one was performed in \cite{Maluf2013,Carvalho} where only one of the operators treated in this paper, that one proportional to $H^{(5)\mu\nu\alpha\beta}_F$, has been considered, however, in this paper we calculate four-derivative terms from a contribution of first order in the LV parameter only, where in \cite{Maluf2013,Carvalho} the second order was studied. In a certain sense our result is more advantageous since corrections of the second order in LV parameters are naturally expected to be suppressed in comparsion with the first-order ones, due to the well-known smallness of LV vectors and tensors, cf. \cite{DataTables}.

We conclude that we succeeded to complete the study of lower-order contributions generated by dimension-5 LV operators that was started in \cite{ourd5}. Possible continuations of this study could consist, first, in development of some its applications in phenomenologically interesting models, second, in studying of perturbative impacts of dimension-6 operators. We plan to perform these studies in our next papers.

{\bf Acknowledgments.}  This work was partially supported by Conselho
Nacional de Desenvolvimento Cient\'{\i}fico e Tecnol\'{o}gico (CNPq). The work by A. Yu. P. has been supported by the
CNPq (project No. 301562/2019-9).

\end{document}